\newcommand{\bm}[1]{\mbox{\boldmath$#1$}}
\documentstyle[preprint,12pt,aps]{revtex}

\input epsf

\title{Phase Transition of a Heisenberg Spin-Glass Model in Three Dimensions }

\author{ F. Matsubara, T. Shirakura$^1$, and S. Endoh }

\address{Department of Applied Physics, Tohoku University, Sendai 980-8579,
Japan \\
$^1$Faculty of Humanities and Social Sciences, Iwate University, 
Morioka 020-8550, Japan}

\date{ \today }

\begin{document}

\maketitle

\begin{abstract}

We study the phase transition of the $\pm J$ Heisenberg model in three 
dimensions. 
Using a dynamical simulation method that removes a drift of the system, 
the existence of the spin-glass (SG) phase at low temperatures is 
suggested. 
The transition temperature is estimated to be $T_{\rm SG} \sim 0.18J$ 
from both equilibrium and off-equilibrium Monte-Carlo simulations.  
Our result contradicts the chirality mechanism of the phase transition 
reported recently by Kawamura which claims that it is not the spins but 
the chiralities of the spins that are ordered in Heisenberg SG systems.

\pacs{75.10.Nr,02.70.Lq,75.40.Cx}

\end{abstract}

%%%%%%%%%%%%%%%%%%%%%%%%%%%%%%%%%%%%%%%%%%%%%%%%%%%%%%%%%%%%%%%%%%%%%%%%%%%%%

It has been believed that the spin-glass (SG) phase is realized in 
three dimensions (3D) for the Ising model \cite{Bhatt,Ogielski} but 
not for the XY and Heisenberg models\cite{Banavar,McMillan1,Olive,Iyota1}. 
Thus, the SG phases observed in experiments were suggested to be realized 
due to magnetic anisotropies\cite{BrayMY,Iyota2}. 
However, it remains puzzling that no sign of Heisenberg-Ising crossover 
has been detected, which is expected if the observed finite temperature 
phase transition is due to a weak magnetic anisotropy. 
Recently, Kawamura and coworkers have proposed a chirality mechanism to 
solve the puzzle\cite{Kawamura2,Kawamura3,Kawamura4,Kawamura5}. 
They considered the chirality described by neighboring three spins. 
It has either positive or negative value, just like the Ising spin. 
Kawamura\cite{Kawamura4} calculated the chirality autocorrelation 
function $C_\chi(t)$ 
and the spin autocorrelation function $C_S(t)$ of 3D Heisenberg SG models 
with and without a weak random magnetic anisotropy $D$ and found that, 
even for $D = 0$, $C_\chi(t)$ exhibits a pronounced aging 
effect reminiscent of the one observed in the mean-field model, 
while $C_S(t)$ exhibits a similar aging effect only when $D \neq 0$. 
He claimed that, 
in the 3D Heisenberg SG model, a chiral-glass (CG) phase transition 
occurs at a finite temperature $T_{\rm CG} \neq 0$, 
but the SG phase is absent. 
He argued that, in real SG magnets, the spin and the chirality are mixed 
due to a weak magnetic anisotropy, and the CG transition is revealed as 
anomalies in experimentally accessible quantities.  
According to this interpretation, the SG phase transition never occurs in 
Heisenberg-like systems, and what was observed in experiments is nothing 
but the CG phase transition. 
This mechanism is quite interesting, because it calls for re-consideration 
of the SG phase transition from both theoretical and experimental 
points of view, because most of SG magnets are rather Heisenberg-like.

%%%%%%%%%%%%%%%%%%%%%%%%%%%%%%%%%%%%%%%%%%%%%%%%%%%%%%%%%%%%%%%%%%%%%%%%%%
In this Letter, we demonstrate that the apparent difference between 
$C_S(t)$ and $C_\chi(t)$ comes from a drift of the system 
and suggest that, contrary to the chirality mechanism, the SG phase 
is realized in the Heisenberg SG model even when the anisotropy is absent. 
We estimate the SG transition temperature of the $\pm J$ Heisenberg model 
as $T_{\rm SG} \sim 0.18J$ 
from both equilibrium and off-equilibrium Monte-Carlo simulations.  
We believe that what was revealed by Kawamura and coworkers is 
nothing but the ordering of the spins themselves. 
The most important point revealed here is that the SG phase will be much 
more stable than what were believed so far 
and re-examination is also necessary for different 
models\cite{Iguchi,Shirakura1,Shirakura2,MSS,KA,Shiomi}.

%%%%%%%%%%%%%%%%%%%%%%%%%%%%%%%%%%%%%%%%%%%%%%%%%%%%%%%%%%\newpage
We start with the $\pm J$ Heisenberg model on a simple cubic lattice of 
$L \times L \times (L+1) ( \equiv N) $ described by 
\begin{eqnarray} 
     H = - \sum_{\langle ij \rangle}J_{ij}\bm{S}_{i}\bm{S}_{j}, 
\end{eqnarray} 
where $\bm{S}_{i}$ is the Heisenberg spin of $|\bm{S}_i| = 1$ 
and $\langle ij \rangle$ runs over all nearest-neighbor pairs. 
The exchange interaction (bond) $J_{ij}$ takes on either $+J$ or $-J$ with 
the same probability of 1/2. 

%%%%%%%%%%%%%%%%%%%%%%%%%%%%%%%%%%%%%%%%%%%%%%%%%%%%%%%%%%%%
First we consider the dynamical properties of the system 
calculating the spin autocorrelation function. 
When one considers the dynamics of the Heisenberg model with 
$D = 0$, one should pay a special attention to a drift of the system, 
i.e., the global rotation of the system.
Although this drift does not affect the static spin correlation and 
becomes negligible in large systems, some data of finite systems 
inevitably suffer from this effect. 
The spin autocorrelation $C_S(t)$ diminishes by this drift, 
whereas the chirality autocorrelation $C_\chi(t)$ does not, because 
the chiralities are invariant under the global rotation of the system. 
Therefore the difference in the behavior between $C_S(t)$ and $C_\chi(t)$ 
might come from this drift. 

%%%%%%%%%%%%%%%%%%%%%%%%%%%%%%%%%%%%%%%%%%%%%%%%%%%%%%%%%%%%%
In order to consider the dynamical properties of an infinitely large system, 
we remove this effect applying the uniform rotation to the system 
so that  
\begin{eqnarray} 
     S = \sum_i| R(t) \bm{S}_{i}(t+t_w) - \bm{S}_{i}(t_w)|^2 
\end{eqnarray} 
becomes minimum, where $R(t)$ is the rotation matrix, and 
$\{\bm{S}_{i}(t_w)\}$ and $\{\bm{S}_{i}(t+t_w)\}$ are the spin 
configurations at times $t_w$ and $t+t_w$, respectively. 
We can successively determine $R(t)$, because $R(t) = E$ at $t = 0$ and it 
changes step by step as $t$ goes by, where $E$ is the unit matrix\cite{Comm0}. 
Hereafter, we call the system described by $\{\bm{S}_i(t)\}$ the drifting 
system and that by $\{ \tilde{\bm{S}}_i(t+t_w) \equiv R(t)\bm{S}_i(t+t_w) \}$ 
the fixed system. 
The spin autocorrelation function $C_S(t_w,t+t_w)$ of the 
fixed system is defined as 
\begin{eqnarray} 
   C_S(t_w,t+t_w) = 
\frac{1}{N}\sum_i[\langle\tilde{\bm{S}}_i(t+t_w)\tilde{\bm{S}}_i(t_w)\rangle], 
\end{eqnarray} 
where $\langle \cdots \rangle$ and [$\cdots$] mean the thermal average and the 
bond distribution average, respectively. 
The chirality autocorrelation function is also defined as 
\begin{eqnarray} 
   C_\chi(t_w,t+t_w) = \frac{1}{3N}\sum_{i\mu}[\langle \chi_{i\mu}(t+t_w) 
                       \chi_{i\mu}(t_w)\rangle],  
\end{eqnarray} 
where $\chi_{i\mu}$ is the chirality at the $i$th site and in the $\mu$th 
direction defined by $\chi_{i\mu} = \bm{S}_{i+\hat{\bm{e}}_\mu} 
\cdot (\bm{S}_i \times \bm{S}_{i-\hat{\bm{e}}_\mu})$ with 
$\hat{\bm{e}}_\mu \; ( \mu = x, y, z)$ being a unit lattice vector 
along the $\mu$ axis. 
Note again that the chirality $\chi_{i\mu}$ is invariant 
under the global spin rotation, and then the chirality autocorrelation 
functions of both drifting and fixed systems are completely the same 
and described by Eq. (4). 
Before applying this method to the present model, we test it in the 
ferromagnetic Heisenberg model, results of which are shown in Fig. 1. 
It is found that the spin autocorrelation function of the fixed system with 
a large $t_w$ approaches to the equilibrium value of the square of the order 
parameter, $\langle (\bm{M}/N)^2 \rangle$, with $\bm{M}$ being the total 
magnetization, whereas that of the drifting system rapidly decays with 
increasing $t$. This result indicates that the effect of the drift 
can be removed properly by the present method.

%%%%%%%%%%%%%%%%%%%%%%%%%%%%%%%%%%%%%%%%%%%%%%%%%%%%%%%%%%%%%%%%%%
We perform the same simulation that was done by Kawamura\cite{Kawamura4} 
using the standard single spin-flip heat-bath Monte-Carlo(MC) method. 
That is, starting with a random initial spin configuration, 
the system is quenched to a working temperature, and after waiting 
$t_w$ MC steps per spin (MCS) the autocorrelation functions are measured 
up to about $2 \times 10^5$ MCS.  
A sample average is taken over about 32 independent bond distributions. 
The results of $C_S(t_w,t+t_w)$ for different 
$t_w$ are presented in Fig. 2 as functions of $t$. 
Contrary to that reported by Kawamura, $C_S(t_w,t+t_w)$ exhibits a quite 
similar aging effect that was found for $C_{\chi}(t_w,t+t_w)$ 
(see Fig. 1 in ref.\cite{Kawamura4}), which indicates that 
the spins and the chiralities possess similar dynamical properties. 
To examine this point in more detail, we plot in Fig. 3 the ratio 
$T(t_w,t+t_w)  \equiv C_\chi/C_S$ for a fixed $t_w = 10^5$ at 
different temperatures. 
At all the temperatures, $T(t_w,t+t_w)$ slowly decreases as $t$ goes by, 
revealing that $C_S$ decays more slowly than $C_{\chi}$. 
This result clearly indicates that, if the ordering of the chiralities 
is realized, then the same is true for the spin. 
We estimate values of the Edward-Anderson order parameters of the spin 
and the chirality, $q_{\rm SG}$ and $q_{\rm CG}$, using the method 
proposed by Parisi {\it et al.}\cite{Parisi} and 
used by Kawamura\cite{Kawamura4}. 
That is, these values are extracted by fitting the data of 
$C_S$ and $C_\chi$ for $t_w = 3 \times 10^5$ to the power-law form of 
\begin{eqnarray}
  C_{S,\chi}(t_w,t+t_w) \sim q_{\rm SG,CG} + \frac{C}{t^{\lambda}}, 
\end{eqnarray}
in the time ranges of $30 \leq t \leq 3 000$ and $50 \leq t \leq 5 000$ . 
The obtained $q_{\rm SG}$ and $q_{\rm CG}$ are plotted as functions of 
temperature in Fig. 4. 
Both quantities have positive, non-vanishing values at low temperatures 
and seem to vanish at almost the same temperature of 
$T \sim 0.18J$\cite{CommOder}. 
This result suggests that, in fact, the phase transitions 
of the spin and the chirality exist at $T \neq 0$ and 
they occur at the same temperature. 
It is not easy, however, to estimate the definite value of the transition 
temperature as well as values of the order parameter 
exponents\cite{KawamuraComm}, 
because the data of $q_{\rm SG}$ around $T = 0.18J$ depend strongly on 
the time range used for extracting them.

%%%%%%%%%%%%%%%%%%%%%%%%%%%%%%%%%%%%%%%%%%%%%%%%%%%%%%%%%%%%%%%%%%
Next we perform a detailed MC simulation in the thermal equilibrium 
to examine the occurrence of the SG phase transition itself. 
We calculate the SG susceptibility $\chi_{\rm SG}$ defined by 
\begin{eqnarray} 
   \chi_{SG} = \frac{1}{N}\sum_{ij}[\langle \bm{S}_i\bm{S}_j\rangle ^2],  
\end{eqnarray}
using the exchange MC algorithm\cite{Hukushima}. 
The sizes of the lattice studied here are $L = 5 \sim 19$. 
Equilibration is checked by monitoring the stability of the results 
against at least two-times longer runs. 
The numbers of the samples are 320 for $L \leq 11$, 96 for $L = 15$, 
and 48 for $L = 19$. The results of $\chi_{\rm SG}$ for different $L$ 
are plotted in Fig. 5. 
At low temperatures, $\chi_{\rm SG}$ exhibits a strong size dependence 
implying the occurrence of the phase transition at $T \neq 0$. 
If the lower critical dimension $d_l$ is less than the lattice dimension, 
$d_l < 3$, and the phase transition really occurs at $T = T_{\rm SG}$, 
the data for different $L$ will be scaled as 
\begin{eqnarray} 
   \chi_{SG} = L^{2-\eta}F(L^{1/\nu}(T-T_{\rm SG})),
\end{eqnarray}
where $\nu$ is the exponent of the correlation length and $\eta$ is the 
exponent which describes the decay of the correlation function at 
$T = T_{\rm SG}$. 
The scaling plots obtained by assuming $T_{\rm SG} \neq 0$ and 
$T_{\rm SG} = 0$ are shown in Fig. 6. 
It should be evident the scaling with $T_{\rm SG} \neq 0$ works 
much better than that with $T_{\rm SG} = 0$, even if the data for 
the smallest size $L = 5$ are ignored in the latter\cite{IyotaComm}. 
The phase transition temperature and the values of the critical exponent 
estimated here are $T_{\rm SG}/J = 0.18 \pm 0.01$, 
$\nu = 0.97 \pm 0.05$ and $\eta = -0.1 \pm 0.1$. 
Note that, in the $\pm J$ model with the random magnetic anisotropy 
$D \neq 0$, the transition temperature was estimated as 
$T_{\rm SG} \sim 0.30J$ for $D = 0.1J$, and $T_{\rm SG} \sim 0.25J$ 
for $D = 0.05J$\cite{Iyota2}. 
Thus our value of $T_{\rm SG} \sim 0.18J$ is consistent with the one 
obtained by extrapolation of those values. 
The value of $\nu \sim 1.0$ for the correlation length also is not strongly 
different from that of $\nu \sim 1.2$ estimated 
in the case of $D \neq 0$\cite{CommD}.

%%%%%%%%%%%%%%%%%%%%%%%%%%%%%%%%%%%%%%%%%%%%%%%%%%%%%%%%%%%%%%%%%%%%%%%%%%%
In summary, all the results presented here seem to support the view 
that the SG phase is realized in the $\pm J$ Heisenberg model and 
the phase transition to this phase occurs at $T_{\rm SG} \sim 0.18J$. 
This view is compatible with our recent studies of the stiffness of the 
system, which suggest that, contrary to previous 
studies\cite{Banavar,McMillan1,Kawamura2,Kawamura3},
the stiffness exponent $\theta_{\rm S}$ of the spin has the positive 
value of $\theta_{\rm S} \sim 0.8$ at $T = 0$\cite{Endoh1} 
and changes its sign at $T \sim 0.19J$\cite{Endoh2}. 
Thus we believe that the SG phase is realized in this model 
even when the anisotropy is absent. 
A major role of the anisotropy will be neither to bring the phase 
transition nor to mix the spin and the chirality, 
but to fix the system and lead anomaly of the 
susceptibility in a finite system\cite{Iyota1}.
One might think that the present suggestion is quite surprising, because 
it disproves the common belief. 
It should be noted again that the occurrence of the ordered phase of the 
chiralities was already suggested by Kawamura et al. and what is suggested  
here is that the spins themselves also order. 
We think the origin of the occurrence of the phase transition 
in the Heisenberg SG model is the ordering of the spins and that the 
ordering of the chiralities happens to be accompanied by it. 
However, further studies are necessary to establish this concept 
as well as to reveal the nature of the phase transition\cite{CommGL}. 
We hope that the present paper stimulates studies of SG systems with the 
continuous spin symmetry.

%%%%%%%%%%%%%%%%%%%%%%%%%%%%%%%%%%%%%%%%%%%%%%%%%%%%%%%%%%%%%%%%%%%%%%%%
\bigskip
\bigskip

The authors would like to thank Professor K. Sasaki,  Dr. T. Nakamura, 
Professor S. Miyake, and Professor H. Takayama for their 
valuable discussions.

%%%%%%%%%%%%%%%%%%%%%%%%%%%%%%%%%%%%%%%%%%%%%%%%%%%%%%%%%%%%%%%%%%%%%%%
%%%%%%%%%%%%%%%%%%%%%%%%%%%%%%%%%%%%%%%%%%%%%%%%%%%%%%%%%%%%%%%%%%%%%%%

%%%%%%%%%%%%%%%%%%%%%%%%%% FIGURES %%%%%%%%%%%%%%%%%%%%%%%%%%%%%%%%%%%%%%%%%%%

\begin{figure}
%%%\vspace*{1cm}
%%%\epsfxsize=1.0\linewidth
%%%\epsfbox{Rotfig1.eps}
%%%\vspace*{-1cm}
\caption{
The spin autocorrelation functions of the ferromagnetic Heisenberg 
model in 3D plotted versus $t$ for a waiting time $t_w = 2 \times 10^4$ 
at temperatures around the Curie temperature of $T_{\rm C}/J \sim 1.45$. 
The filled and open symbols are data from the fixed and the drifting 
systems, respectively. The arrows indicate the equilibrium values of 
the square of the magnetization $\langle (\bm{M}/N)^2 \rangle$.  
}
\label{fig:1}
\end{figure}
\begin{figure}
%%%\vspace*{1cm}
%%%\epsfxsize=1.0\linewidth
%%%\epsfbox{Rotfig2.eps}
%%%\vspace*{-1cm}
\caption{
The spin autocorrelation functions of the $\pm J$ Heisenberg model 
plotted versus $t$ for different waiting time $t_w$. 
The data are averaged over 32 samples of the lattice size $L = 15$. 
}
\label{fig:2}
\end{figure}
\begin{figure}
%%%\vspace*{1cm}
%%%\epsfxsize=1.0\linewidth
%%%\epsfbox{Rotfig3.eps}
%%%\vspace*{-1cm}
\caption{
The ratio of the chirality to the spin autocorrelation functions 
$T(t_w,t+t_w) \equiv C_{\chi}/C_S$ of the $\pm J$ Heisenberg model 
plotted versus $t$. 
The data are averaged over 32 samples of the lattice size $L = 15$. 
}
\label{fig:3}
\end{figure}
\begin{figure}
%%%\vspace*{1cm}
%%%\epsfxsize=1.0\linewidth
%%%\epsfbox{Rotfig4.eps}
%%%\vspace*{-1cm}
\caption{
Temperature dependences of the Edwards-Anderson order parameters of the 
spin and the chirality, $q_{\rm SG}$ and $q_{\rm CG}$, of the $\pm J$ 
Heisenberg model. Open and solid symbols indicate those extracted 
in the time ranges of $30 \leq t \leq 3000$ and $50 \leq t \leq 5000$, 
respectively. The data are averaged over 32 - 200 samples. 
Lines with $\beta = 0.5$ for $q_{\rm SG}$ and $\beta_{\rm CG} = 
1.0$ for $q_{\rm CG}$ are guides to the eye. 
}
\label{fig:4}
\end{figure}
\begin{figure}
%%%\vspace*{1cm}
%%%\epsfxsize=1.0\linewidth
%%%\epsfbox{Rotfig5.eps}
%%%\vspace*{-1cm}
\caption{
Temperature dependences of the spin-glass susceptibility $\chi_{\rm SG}$ 
of the $\pm J$ Heisenberg model in 3D for different sizes of the lattice. 
}
\label{fig:5}
\end{figure}
\begin{figure}
%%%\vspace*{1cm}
%%%\epsfxsize=1.0\linewidth
%%%\epsfbox{Rotfig6.eps}
%%%\vspace*{-1cm}
\caption{
Typical examples of the finite size scaling plot for (a) $T_{\rm SG} \neq 0$ 
and (b) $T_{\rm SG} = 0$.
}
\label{fig:6}
\end{figure}

%%%%%%%%%%%%%%%%%%%%%%%%%%%%%%%%%%%%%%%%%%%%%%%%%%%%%%%%%%%%%%%%%%%%%%%%%%%%%


\begin{references}
%
\bibitem{Bhatt} R. N. Bhatt and A. P. Young, Phys. Rev. Lett. {\bf 54}, 924 
           (1985).
%
\bibitem{Ogielski} A. T. Ogielski and I. Morgenstern, Phys. Rev. Lett. 
           {\bf 54}, 928 (1985); 
           A. T. Ogielski, Phys. Rev. B {\bf 32}, 7384 (1985).
%
\bibitem{Banavar} J. R. Banavar and M. Cieplak, Phys. Rev. Lett. 
{\bf 48}, 832 (1982). 
%
\bibitem{McMillan1} W. L. McMillan, Phys. Rev. B {\bf 31}, 342 (1984).
%
\bibitem{Olive} J. A. Olive, A. P. Young, and D. Sherrington, 
           Phys. Rev. B {\bf 34}, 6341 (1986). 
%
\bibitem{Iyota1} F. Matsubara, T. Iyota and S. Inawashiro, 
          J. Phys. Soc. Jpn. {\bf 60}, 4022 (1991).
%
\bibitem{BrayMY} A. J. Bray, M. A. Moore and A. P. Young, 
          Phys. Rev. Lett. {\bf 56}, 2641 (1986).
%
\bibitem{Iyota2} F. Matsubara, T. Iyota, and S. Inawashiro, 
          Phys. Rev. Lett. {\bf 67}, 1458 (1991).
%
\bibitem{Kawamura2} H. Kawamura, Phys. Rev. Lett. {\bf 68}, 3785 (1992). 
%
\bibitem{Kawamura3} H. Kawamura, J. Phys. Soc. Jpn. {\bf 64}, 26 (1995). 
%
\bibitem{Kawamura4} H. Kawamura, Phys. Rev. Lett. {\bf 80}, 5421 (1998). 
%
\bibitem{Kawamura5} K. Hukushima and H. Kawamura, Phys. Rev. E {\bf 61}, 
                    R1008 (2000). 
%
\bibitem{Iguchi} F. Matsubara and M. Iguchi, 
        Phys. Rev. Lett. {\bf 68}, 3781 (1992).
%
\bibitem{Shirakura1} T. Shirakura and F. Matsubara,
         J. Phys. Soc. Jpn. {\bf 65}, 3138 (1996).
%
\bibitem{Shirakura2} T. Shirakura and F. Matsubara, 
         Phys. Rev. Lett. {\bf 79}, 2887 (1997).
%
\bibitem{MSS} F. Matsubara, T. Shirakura, and M. Shiomi, 
        Phys. Rev. B {\bf 58}, R11821 (1998). 
%
\bibitem{KA} J. M. Kosterlitz and N. Akino,
         Phys. Rev. Lett. {\bf 82}, 4094 (1999).
%
\bibitem{Shiomi} M. Shiomi, F. Matsubara, and T. Shirakura, 
         J. Phys. Soc. Jpn. {\bf 69}, 2798 (2000).
%
\bibitem{Comm0} It is well known that any rotation matrix is described 
      by $R(t) = \exp(D)$ with $D$ being an antisymmetric matrix. 
      If $R(t)$ is given as $R(t) = \exp(D_t)$, then $R(t+1)$ can be readily 
      obtained by putting $R(t+1) = \exp(\delta D)\exp(D_t)$ and 
      using Newton-Raphson method. 
%
\bibitem{Parisi} G. Parisi, F. Ricci-Tersenghi, and J. J. Ruiz-Lorenzo, 
        J. Phys. A {\bf 29}, 7943 (1996).
%
\bibitem{CommOder} Since $q_{\rm SG}$ and $q_{\rm CG}$ should be positive 
       or 0, the extrapolation function of eq. (5) becomes meaningless 
       for $T \geq 0.19J$ for $q_{\rm CG}$ and $T \geq 0.20J$ 
       for $q_{\rm SG}$. This fact implys that $T_{\rm CG} < 0.19J$ 
       and $T_{\rm SG} < 0.20J$. 
%
\bibitem{KawamuraComm} Kawamura\cite{Kawamura4} estimated the order-parameter 
       exponent of $\beta_{\rm CG} \sim 1.1$ for $q_{\rm CG}$ and 
       suggested that the universality class of the Heisenberg-like SG 
       magnets might differ from that of the standard Ising SG which is 
       characterized by the exponent of $\beta \sim 0.5$ for $q_{\rm SG}$. 
%
\bibitem{Hukushima} K. Hukushima and K. Nemoto, 
       J. Phys. Soc. Jpn. {\bf 65}, 1604 (1996).
%
\bibitem{IyotaComm} In the previous scaling analysis, $T_{\rm SG} = 0$ was 
     estimated using the data for lattices of $L = 7 - 15$\cite{Iyota2}. 
     Here, we re-examine the phase transition using the data for wider 
     temperature ranges and adding the data of the bigger lattice of $L = 19$.
%
\bibitem{CommD} The value of $\eta \sim -0.1$ seems to deviate markedly 
     from that of $\eta \sim 0.6$ for the case of $D = 0.05$\cite{Iyota2}. 
     However, we note that $\eta$ depends strongly on the estimation of 
     $T_{\rm SG}$. In that case, if we takes the lower bound of 
     $T_{\rm SG} = 0.24J$, we get $\eta \sim 0.1$. 
%
\bibitem{Endoh1} F. Matsubara, S. Endoh, and T. Shirakura, 
        J. Phys. Soc. Jpn. {\bf 69}, 1927 (2000). 
%
\bibitem{Endoh2} S. Endoh, F. Matsubara, and T. Shirakura; in preparation.
%
\bibitem{CommGL} It was reported that each of the spin Binder parameter 
        $g_{\rm SG}$ and the chirality Binder parameter $g_{\rm CG}$ 
        exhibits a strange temperature dependence\cite{Kawamura5}. 
        Further studies are necessary to make clear the relationship 
        between the phase transition and the behaviors of these 
        Binder parameters. 
%
\end{references}
\end{document}